\documentclass[twocolumn,showpacs,preprintnumbers,amsmath,amssymb]{jpsj3}
\usepackage{dcolumn}

\def\runauthor{Online-Journal Subcommittee}

\title{%
 Dynamics of Superflow by Mesoscopic Condensate
}

\author{%
Shun-ichiro Koh\thanks{E-mail address: koh@kochi-u.ac.jp}
}

\inst{%
 Physics Division, Faculty of Education, Kochi University  \\
        Akebono-cho, 2-5-1, Kochi, 780, Japan 
}

\recdate{\today}

\abst{%
The  shear viscosity $\eta $ of a quantum liquid in the vicinity of  
 $T_{\lambda}$ is examined. In liquid helium 4 above $T_{\lambda}$
 ($T_{\lambda}<T<3.7K$), under a strong effect of Bose statistics, 
 the coherent many-body wave function grows to an 
 intermediate size between a macroscopic level and a microscopic one. 
 These wave functions are qualitatively different from thermal fluctuation, and  
 manifest themselves in the gradual decrease in shear viscosity above $T_{\lambda}$. 
 To formulate this phenomenon, we combine the correlation function 
 with fluid dynamics.  Applying the Kramers-Kronig relation to the generalized Poiseuille's 
 formula for capillary flow, we perform a perturbation calculation  of 
  the reciprocal $1/\eta $ with respect to the particle interaction, and examine how
the growth of coherent wave functions gradually decreases shear viscosity.
 Comparing with the experimentally determined $\eta (T)$, $\hat 
 {\rho¥}_s(T)/\rho¥$ of such a mesoscopic condensate is estimated to   
 reach $10^{-5}$ just above $T_{\lambda}$. We examine the effect 
 of condensate size on the stability of such a superflow, and 
touch upon the superflow in porous media.
}

\kword{%
Superfluidity, Capillary flow, Liquid helium 4, Shear viscosity, 
Mesoscopic condensate
}

\begin{document}
\maketitle

\section{Introduction}

Superfluidity was first discovered in a macroscopic frictionless flow of liquid helium 4
through a capillary or a narrow slit  \cite {kap}. Since then,
 the range of studies of superfluidity has expanded. 
Since the late 1970's, when people had an impression that our understanding of 
superflow in bulk liquid helium 4 was essentially completed \cite {lan}, 
experimenters have turned their interest to superflow in 
 more exotic structures such as porous media  \cite {rev}.
These materials allow us to study new aspects of superfluidity, so far 
inaccessible in the bulk liquid helium 4. As is often in physics, 
however, the original capillary flow experiment that first 
demonstrated superfluidity is not the most conceptually clear-cut 
demonstration of it. Capillary flow is well known but not completely 
understood  even at present  \cite {vis}. One can think of two reasons: 
(1) Capillary flow is a  nonequilibrium phenomenon  \cite {tex} \cite {kub}, 
which exhibits subtle features that do not exist in the equilibrium 
manifestation of superfluidity  such as nonclassical rotational 
properties. (2) In superfluid helium 4, many features associated with 
Bose statistics are masked by the strongly interacting nature of the 
liquid. The formulation of the viscosity of a classical liquid, especially 
 from the structural viewpoint  \cite {fre}, has been a difficult 
problem, because complex motions inherent in the liquid do not 
allow us to make a simple microscopic formulation.  

\begin{figure}
\includegraphics [scale=0.56]{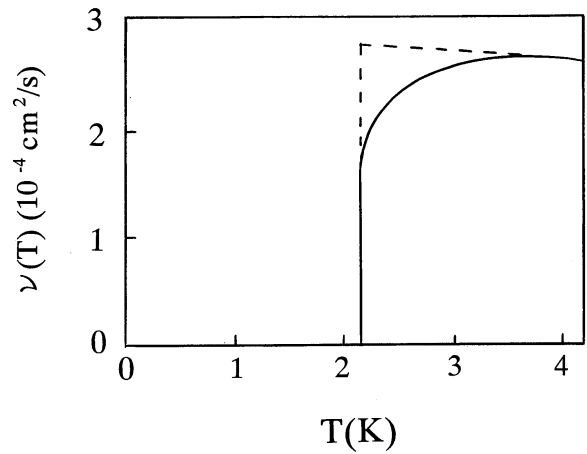}
\caption{\label{fig:epsart} 
   The kinematic viscosity $\nu (T)$ (a solid curve) of a bulk liquid 
   helium 4. }
\end{figure}

In this paper, the capillary flow above $T_{\lambda}$ is revisited. 
The two-fluid model, which is the basis for understanding liquid helium 4, separates the system into 
the normal and superfluid parts from  the beginning, and normally assumes that 
the latter abruptly appears at $T_{\lambda}$.  The success of the 
two-fluid model instilled into us the notion that superflow appears at $T_{\lambda}$ in 
a mathematically discontinuous manner, and that all anomalous 
properties above $T_{\lambda}$ are due to thermal fluctuation. 
(In fact,  in liquid helium 4 at $|T/T_{\lambda}-1|<10^{-2}$, 
the sharp increase in specific heat or the softening of sound 
propagation arises from thermal fluctuation.)   
Figure 1 shows the temperature dependence of the kinematic 
viscosity $\nu (T)=\eta (T)/\rho$ of the capillary flow of liquid helium 4
\cite {men} \cite {zin} \cite {bar}. 
In Fig.1, when cooling the system, the kinematic viscosity $\nu (T)$ does not abruptly 
drops to zero at $T_{\lambda}$ like a dotted line, but it gradually 
decreases in $T_{\lambda}<T<3.7K$, and finally drops to zero at  $T_{\lambda}$.  
Similarly to the anomalies at $|T/T_{\lambda}-1|<10^{-2}$, 
this gradual decrease in $\nu (T)$ in $T_{\lambda}<T<3.7K$ has been 
tacitly attributed to thermal fluctuation (except in ref.8).
Actually, behind the two-fluid model, there exists a more basic 
assumption in statistical physics: the infinite-volume limit. 
To clearly define the phase transition, 
this limit  eliminates the intermediate-sized order, in which     
``intermediate size'' is equivalent to ``microscopic one''.  
For the mechanical properties,  however, in contrast to the 
thermodynamic ones, one cannot ignore the boundary condition of the 
system.  In particular, in this system, since the size of the coherent wave 
function and that of the capillary are not so different at low 
temperatures, we cannot naively 
use the $V\rightarrow \infty$ limit.  Rather, the magnitude of phenomena
masked by this limit should be estimated by experiments. (In this sense, the 
 low-temperature phenomenon with quantum coherence is essentially a mesoscopic one.)

Physically, the interpretation of $\nu (T)$ in $T_{\lambda}<T<3.7K$ on the basis of 
 thermal fluctuation is questionable for the following reasons.  
(1) The range of temperature of 1.5 K from $T_{\lambda}$ to 3.7 K is too large for 
the thermal fluctuation \cite {ide}. For the fluctuation of 
temperature $\Delta T$, one knows the formula $\langle (\Delta T)^2\rangle =k_BT^2/C_V$, 
in which $C_V$ is the heat capacity of a fluctuating small region. 
As mentioned in ref.8, the $C_V$ in the case of $\Delta T= 1.5 K$ is 
about $2.8 \times 10^{-23}J/K$, which implies that the fluctuating region 
would have to be as small as one atom diameter. 
(2) Thermal fluctuation gives rise to the short-lived and randomly 
oriented wave function obeying Bose statistics.  In contrast, 
 for the one-directional flow to show very small shear viscosity \cite {sma},  
{\it a long-lived and coherent translational motion of particles along a specific 
direction \/} is necessary, which is qualitatively different from thermal fluctuation.
(In this sense, the fluctuation-dissipation theorem is not applicable to the 
nondissipative response  such as the decrease in viscosity \cite {inc}.)
Rather, the gradual fall of $\nu (T)$ in $T_{\lambda}<T<3.7K$ is 
likely to represent {\it the orderly redistribution of 
 particle momenta due to Bose statistics \/}, an advance sign of the state 
 below $T_{\lambda}$  \cite {men}.
 
In the Bose system just above $T_{\lambda}$, particles experience a strong 
effect of Bose statistics, and the coherent wave function grows to a large 
but not yet macroscopic size. 
Since these wave functions are negligible at $V\rightarrow \infty$, 
they do not affect the definition of phase transition at $T_{\lambda}$.
However, in contrast to thermal fluctuation, these intermediate-sized 
coherent wave functions have a possibility of leading to a coherent translational motion of 
particles within a mesoscopic distance \cite {koh}. At $T_{\lambda}<T<3.7 K$, there is no 
macroscopic condensate connecting the two ends of the capillary, 
hence no macroscopic frictionless transport occurs. Rather, the intermediate-sized 
coherent wave function is likely to reduce the $\nu (T)$ of capillary 
flow. We will call such a wave function 
{\it mesoscopic condensate \/} not in the thermodynamic sense but in the mechanical sense.
 Figure.1 must include valuable information on the mesoscopic condensate. 
Our problem is to estimate it by experiment and 
compare it with the microscopic model.  For this purpose, we must take 
a somewhat different approach to superflow.

(1) The shear viscosity of a quantum liquid has been studied by 
applying the kinetic theory of gases to phonons and rotons  \cite {kha}. 
However, for our purpose, this method is not applicable. 
 Well below $T_{\lambda}$, various excitations of liquid helium 4 are strictly 
 suppressed except for phonons and rotons.  Hence, they are normally regarded as 
a weakly interacting dilute Bose gas, although 
the excitation in liquid.  For $\nu (T)$ above $T_{\lambda}$, however, the 
 dilute-gas picture has no basis, because the basis for this 
 picture, the macroscopic condensate,  has not yet developed. 
Rather, the strongly interacting excitations in the liquid 
 determine the $\nu$ above $T_{\lambda}$. The effect of Bose 
statistics on such excitations is worth studying. 

\begin{figure}
\includegraphics [scale=0.4]{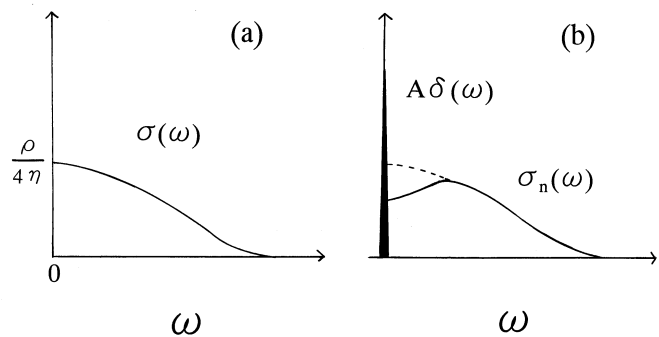}
\caption{\label{fig:epsart} The change of the conductivity spectrum 
$\sigma (\omega )$ from (a) in a classical fluid to (b) in a superfluid. 
   }
\end{figure}

In \S 2, we will develop an alternative method combining the 
correlation function with fluid dynamics \cite {kad}.  
The distribution of flow velocity in a capillary is described by Poiseuille's formula
\begin{equation}
 v_x(r)=\frac{d^2-r^2}{4\eta¥}¥\frac{\Delta P}{L¥}¥,
	\label{¥}
\end{equation}¥
 where $d$ and $L$ are the radius and length of the capillary, respectively, and 
$\Delta P$ is the pressure between the two ends of the capillary.  
The simplest method of measuring $\eta $ is to set the capillary to stand 
vertically at gravity $g$. The level $h(t)$ of the liquid decreases as 
$h(t)=h(0)\exp (-\alpha t)$, where $\alpha$ includes $\eta$ as $\rho 
gd^4/(\eta L)$ (see Appendix. A).
Hence, $\Delta P(t)$ also decreases as $\Delta P(t)=\Delta 
P(0)\exp (-\alpha t)$ in eq.(1). In $T_{\lambda}<T<3.7K$, the curve of $h(t)$ shows 
an anomalously small shear viscosity. Although it shows no sign of 
frictionless transport, the frequency spectrum of the quantities in eq.(1) 
reflects the mesoscopic dynamics in the flow as well as the macroscopic one, and 
therefore, mesoscopic superflow must exhibit a characteristic spectrum in it. 
 On the axis of the capillary ($r=0$), multiplying both sides of eq.(1) by the density $\rho $, we define 
the massflow density $j(t)=\rho v_x(0,t)$, and the conductivity $\sigma =\rho /(4\eta¥)$.
Under an oscillating $P (\omega )\exp (i\omega t)$, the spectrum 
$j(\omega)$ and $\sigma (\omega )$ (Fig.2(a)) satisfy 
\begin{equation}
 \mbox{\boldmath $j$}(\omega)=-\sigma (\omega)¥¥d^2 \frac{\Delta \mbox{\boldmath $P$}(\omega)}{L¥}.
	\label{¥}
\end{equation}¥

In the capillary flow below $T_{\lambda}$ or in the vicinity of $T_{\lambda}$, the 
superfluid and normal-fluid parts flow without any transfer of momentum to each 
other. Accordingly, $\sigma (\omega)$  splits into a sharp peak $A\delta(\omega)$ at 
$\omega =0$ and a continuous spectrum $\sigma _n(\omega)$ as in Fig.2(b); 
\begin{equation}
 \mbox{\boldmath $j$}(\omega)=-\left[\sigma _n(\omega)+A\delta(\omega)\right]¥¥d^2 
 \frac{\Delta \mbox{\boldmath $P$}(\omega)}{L¥}.
	\label{¥}
\end{equation}¥
In \S 2, eqs.(2) and (3) will be embedded into the general linear-response 
relation including both the dissipative and nondissipative (dispersive) responses. 
Applying the Kramers-Kronig relation to them, we will derive a formula of 
$\nu (T)=1/[4\sigma (T)]$ (eq.(11)) in the vicinity of $T_{\lambda}$, and 
quantitatively estimate the effect of Bose statistics in Fig. 1. 
In \S 2.3, the role of Bose statistics in the fall of 
shear viscosity will be discussed using the Maxwell relation.

(2) In the microscopic theory of shear viscosity $\eta $, one normally starts from 
the two-time correlation function of the tensor 
$J_{xy}(t)=-\sum_{i}(p_{i,x}p_{i,y}/m¥)$, and perform a perturbation 
calculation of $\langle J_{xy}(0)J_{xy}(t)\rangle$ with respect to 
the particle interaction  \cite {tex} \cite {kub}. This method has the 
following difficulties in solving our problem.  
In the weak-coupling system (a gas, or a simple liquid such as 
liquid helium 4), the particle interaction $U$ normally enhances the 
relaxation of particles to local equilibrium positions in the flow, thereby 
decreasing $\eta $ \cite {stro}. If we would try to formulate this 
property using the  Kubo formula for $\eta $, we must derive the decrease in $\eta$ from 
the increase in $U$. Since $\eta $ and $U$ changes in opposite directions,  
there must be a delicate cancellation of higher-order terms in the 
perturbation expansion of $\eta $. 
In contrast, when we start from Poiseuille's formula, in which 
 $\eta$ appears in the denominator of the linear-response coefficient as in  eq.(1), 
 we apply the Kubo formula not to $\eta $ but to the {\it reciprocal \/} $1/\eta $ in eq.(2).
In the perturbation expansion of $1/\eta$ with respect to $U$, 
{\it the increase in $U$ naturally leads to the increase in $1/\eta$,  
and therefore to the decrease in $\eta$ \/}.  In this case, one need not expect the 
cancellation, and the effect of $U$ on $\eta $ is 
simply built in from the beginning. 

(3) We will focus on the redistribution of particle momenta in the vicinity of  
$T_{\lambda}$. In \S 3, following ref.14, we assume the Bose system without 
the macroscopic condensate as a non-perturbed state, 
and perform a perturbation calculation of $1/\eta $. 
 By taking characteristic diagrams reflecting Bose statistics in the expansion, we will examine how 
 the formation of a larger coherent wave function  gradually decreases shear viscosity.
  Specifically, combining in \S 3 the phenomenological relation 
eq.(3) with the microscopic calculation, we will derive the formula of $\sigma(\omega)$ (eq.(28))
illustrated in Fig.2(b) , and derive the suppression of $\nu $. 

(4) Compared with macroscopic condensate, the mesoscopic condensate 
shows different responses in the oscillation experiments.
After defining the damping angular frequency (eq.(24)), 
we will examine the stability of superflow to the oscillation, and 
estimate the effect of condensate size on the stability (eq.(31)).  
From this viewpoint, we will briefly discuss the response of the superflow in porous media 
to torsional oscillation or ultrasound in \S 4.

Lastly, we will discuss these results in a wider context in \S 5.

\section{Shear Viscosity in Capillary Flow}

\subsection{Formalism}
In classical fluids, the capillary flow is a typical dissipative process.
To describe both dissipative and nondissipative processes, let us 
generalize the conductivity spectrum $\sigma (\omega)$ in eq.(2) to the complex number 
\begin{equation}
 \mbox{\boldmath $j$}(\omega)=-\left[\sigma_1(\omega)+i\sigma_2(\omega)\right]¥d^2 
 \frac{\Delta \mbox{\boldmath $P$}(\omega)}{L¥}.
	\label{¥}
\end{equation}¥
The spectrum of $\Delta P(t)=\Delta P(0)\exp (-\alpha t)$ is given by
 $\Delta P(\omega)=\Delta P(0)\pi ^{-1}(\omega ^2+\alpha ^2)^{-1/2}$, 
 which has a peak at $\omega =0$.  We define the half width $\omega _f$ of 
 the peak as  $\omega _f=\sqrt{3}\alpha¥$. In eq.(4),
$\sigma _1(\omega)$ must satisfy the following sum rule \cite {kub} 
including the conserved quantity $f(d)$  (see Appendix. B) 
\begin{equation} 
   \frac{1}{\pi ¥}¥\int_{0}^{\infty¥}\sigma _1(\omega)d\omega¥=f(d)¥.
	\label{¥}
\end{equation}¥
The average conductivity $\sigma$ is defined as
$\sigma=\omega _f^{-1}\int_{0}^{\omega _f¥}¥\sigma _1(\omega)d\omega$,
hence giving the shear viscosity $\eta =\rho /(4\sigma¥)$. 

To incorporate eq.(4) into the linear-response theory, instead of $\Delta P(\omega)$, 
we will use {\it the velocity \/} $v_0(\omega)$, which satisfies the equation 
of motion $\rho dv_0(t)/d t¥= -\Delta P(t)/L¥$.
Since $v_0(\omega)$ and $j(\omega)$ constitute the perturbation 
energy $\int ¥j(\omega)v_0(\omega)d\omega$, we can regard $v_0(\omega)$ 
as the external field in eq.(4). We rewrite eq.(4) as 
\begin{equation}
 \mbox{\boldmath $j$}(\omega)=\rho\left[-\omega\sigma_2(\omega)+i\omega\sigma_1(\omega)\right]¥¥d^2
                                        \mbox{\boldmath $v$}_0(\omega),
           \label{¥}
\end{equation}¥
and regard $\rho\left[-\omega\sigma_2(\omega)+i\omega\sigma_1(\omega)\right]¥¥d^2$ 
as the coarse-grained form of generalized susceptibility. 
In this formula, the real part $-\omega\sigma_2(\omega)$  
corresponds to  the non-dissipative process, and the imaginary part 
$\omega\sigma_1(\omega)$ to the dissipative one. In liquid, the former corresponds 
to the Couette flow of viscous liquid in a rotating bucket. 
When the whole fluid is in the uniform rotation, no internal friction 
occurs \cite {dis}, and it rotates like a rigid body 
as $\mbox{\boldmath $v$}_d( \mbox{\boldmath $r$})= \mbox{\boldmath 
$\Omega$}\times \mbox{\boldmath $r$}$ ($\Omega$ is the angular velocity). 
The latter corresponds to the  capillary flow.  
They are connected to each other by the Kramers-Kronig relation \cite {fer} as 
\begin{equation}
   \sigma_1(\omega')=\frac{2}{\pi}¥\int_{0}^{\infty¥}d\omega\frac{\omega\sigma_2(\omega)}{\omega^2-\omega'^2¥}¥.
	\label{¥}
\end{equation}¥

 Using the response of the rotating liquid as the real part $\omega\sigma_2(\omega)$, 
 we can obtain the conductivity of the capillary flow $\sigma_1(\omega)$ by eq.(7). 
$\omega\sigma_2(\omega)$ is derived from the susceptibility $\chi_{\mu\nu}(q,\omega )$ 
(the real part of the generalized one) \cite {noz}. In general, the  $\chi_{\mu\nu}(q,\omega )$ 
is decomposed into the longitudinal and transverse parts,  
$\chi_{\mu\nu}(q,\omega )=\chi^L(q,\omega) q_{\mu}q_{\nu}/q^2¥     
+\chi^T(q,\omega)\left(\delta_{\mu\nu}-q_{\mu}q_{\nu}/q^2\right)¥ $, 
in which $\omega\sigma_2(\omega)$  belongs to the transverse part.
(In the case of the rotation, the effect of the wall of the bucket propagates along 
the radial direction, which is perpendicular to that of particle 
motion.)  Hence, we rewrite eq.(7) as
\begin{equation}
   \rho d^2\sigma_{1}(\omega')= -\frac{2}{\pi}¥¥
         \int_{0}^{\infty¥}d\omega\frac{\lim _{q\rightarrow 0}\chi^T(q,\omega)}{\omega^2-\omega'^2¥}¥.
	\label{¥}
\end{equation}¥
 In the normal-fluid phase, because of $\chi ^T(q,\omega)=\chi ^L(q,\omega)$ for  small $q$, 
 $\chi^T(q,\omega)$ in eq.(8) is replaced by the longitudinal part $\chi ^L(q,\omega)$. 
Such a modified formula with the aid of $\rho =\chi ^L(q,0)$ defines 
the normal fluid conductivity $\sigma_{1n}(\omega)$ (see Appendix. B).

As $T\rightarrow T_{\lambda}$, owing to Bose statistics, the 
energy of the low-lying transverse excitation increases (see \S 2.3). 
It destroys the balance between the transverse and the longitudinal excitations,  
hence $\chi^L(q,\omega)-\chi^T(q,\omega)¥\ne 0$ at $\omega \rightarrow 0$. 
Instead of $\sigma_{1n}(\omega)$ defined by $\chi^L(q,\omega)$, one must 
go back to the original definition of 
eq.(8), and rewrite it using $\sigma_{1n}(\omega)$ as 
 \begin{eqnarray}
  \lefteqn{\sigma_1(\omega')=\sigma_{1n}(\omega')} \nonumber \\ 
         &&+\frac{2}{\rho\pi d^2¥}¥\int_{0}^{\infty¥}d\omega
            \frac{\lim _{q\rightarrow 0}[\chi^L(q,\omega)-\chi^T(q,\omega)¥]}{\omega^2-\omega'^2¥}¥.
	\label{¥}
\end{eqnarray}¥

The integral over $\omega$ in Eq.(9) leads to a sharp peak at $\omega '=0$ \cite {fer}
with the aid of Hilbert transformation as
\begin{equation}
  \int_{0}^{\infty¥}¥\frac{d\omega}{\omega^2-\omega'^2¥}¥=\frac{\pi ^2}{2¥}¥\delta(\omega ') .
	\label{¥}
\end{equation}¥
Accordingly, a sharp peak $\lim _{q\rightarrow 0}[\chi^L(q,0)-\chi^T(q,0)]\delta(\omega)$
appears in $\sigma_1(\omega)$ of eq.(9) as required in eq.(3), 
 where $\lim _{q\rightarrow 0}[\chi^L(q,0)-\chi^T(q,0)¥]$ corresponds
to the mesoscopic superfluid density $\hat {\rho¥}_s(T)$. 
 Using  eq.(10) in eq.(9), and using the result in $\sigma=\omega _f^{-1}\int_{0}^{\omega _f¥}¥\sigma 
(\omega)d\omega$, we take out the sharp peak from this integral, and obtain the 
kinematic viscosity $\nu (T)=1/(4\sigma)$ as  \cite {del}
\begin{equation}
  \nu (T)=¥\frac{\nu _n}{1
              +\displaystyle{\frac{2\pi}{d^2\omega _f ¥} \frac{\hat 
              {\rho _s}(T)}{\rho}}\nu _n¥¥}¥, 
	\label{¥}
\end{equation}¥
where $\nu _n$ ($=1/[4\sigma _n]$) is $\nu$ of the classical liquid. 

\subsection{Comparison with the experiment}
\begin{figure}
\includegraphics [scale=0.4]{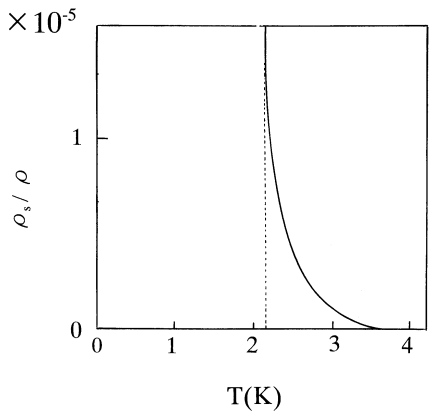}
\caption{\label{fig:epsart} 
 $\hat {\rho¥}_s(T)/\rho$ obtained by eq.(12) using $\nu (T)$ in 
  Fig.1.  }
\end{figure} 
Equation (11) is a general formula depending on no specific model of liquids, 
and we can compare it to $\nu (T)$ shown in Fig.1 \cite {bar}. 
When we view the fall of $\nu (T)$ in Fig.1 as a manifestation of a 
mesoscopic condensate, we can estimate its $\hat{\rho _s}/\rho$ using eq.(11) as follows.
We assume $\nu _n$ in eq.(11) to be $\nu $ at $T_{on}$=3.7K in Fig. 1, 
and derive $\omega _f$ in eq.(11) from the time dependence of the 
 level of a liquid $h(t)=h(0)exp(-\alpha t)$ as follows. 
 The $\alpha$ of liquid helium 3 was measured at 1.105 K 
as $5\times 10^{-4}/s$ in Fig.2 of ref.9. 
At 3.7 K, it approximately changes to $3\times 10^{-4}/s$ with a 
change of $\rho /\eta$ in $\alpha \propto \rho gd^4/(\eta L)$. 
For liquid heliums 3 and 4 near 3.7 K, the factor $\rho /\eta$ is nearly the same. 
Hence, we estimate that liquid helium 4 at 3.7 K has $\alpha = 
3\times 10^{-4}/s$ for a typical size of the capillary, and 
$\omega _f =\sqrt{3}¥\alpha =5\times 10^{-4}rad/s$ (a small value).
 With the above $\omega _f$ and a typical capillary radius $d=10^{-2}cm$, 
 we estimate $\hat {\rho¥}_s(T)/\rho¥$ using $\nu (T)$ of Fig.1 as 
\begin{equation}
 \frac{2\pi}{d^2\omega _f¥}¥\frac{\hat {\rho¥}_s(T)}{ \rho¥}¥ =\nu (T)^{-1}-\nu (T_{on})^{-1},
	\label{¥}
\end{equation}¥ 
and obtain Fig. 3. 
Just above $T_{\lambda}$, $\hat {\rho¥}_s(T)/\rho¥$ reaches $1.6\times 10^{-5} $.

Compared with the nondissipative process such as the rotation of liquid, 
the intrinsically dissipative process such as the capillary flow depends 
on some experimental parameters such as $\omega _f$ and $d$ as in eq.(11). 
In the rotating bucket experiment, the change in the moment of inertia 
$I_z$ simply obeys $I_z(T)=I_z^{cl}[1-\hat {\rho¥}_s(T)/\rho¥]$. 
 In the rotation experiment by Hess and Fairbank \cite {hes}, the 
 moment of inertia $I_z$ above $T_{\lambda}$ is slightly smaller than the  normal 
 phase value $I_z^{cl}$.  Using these  currently available data,  
 $\hat{\rho _s}(T_{\lambda}+0.03K)/\rho\cong 8\times 10^{-5}$, and  
 $\hat{\rho _s}(T_{\lambda}+0.28K)/\rho\cong 3\times 10^{-5}$ in ref.14.
  Despite the subtleness inherent in the dissipative process, $\hat 
  {\rho}_s(T)/\rho $ in Fig.3 has the same order of 
 magnitude as $\hat  {\rho}_s(T)/\rho $ independently obtained using $I_z(T)$. 
Just above $T_{\lambda}$, about $1/10^5$ of all helium 4 atoms participate in the mesoscopic 
condensate.  They are negligible in thermodynamic quantities, but 
manifest themselves in the mechanical response such as shear viscosity. 
In eq.(11), the effect of $\hat {\rho}_s(T)$ is amplified by the small $\omega _f$.

\begin{figure}
\includegraphics [scale=0.5]{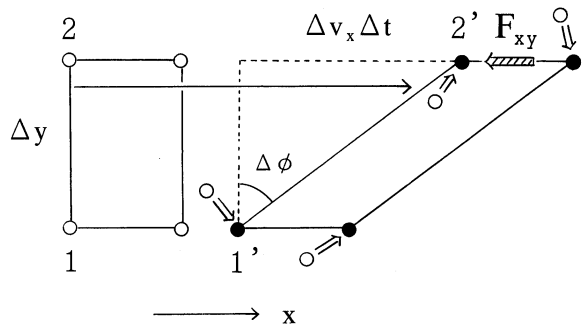}
\caption{\label{fig:epsart} Deformation of a small rectangular 
element in the flow.  The shear stress $F_{xy}$ is expressed as
    $\eta \partial v_x/\partial y$. }
\end{figure}

\subsection{Physical picture in coordinate space}
(1) Let us discuss the physical mechanism behind eq.(11).
 Figure 4 shows a small rectangular element in the flow, which initially 
 deforms macroscopically, then relaxes to a part of the stationary flow 
 with a relaxation time $\tau _{M}$. 
 For $\eta$ of the classical liquid, Maxwell obtained a simple 
 formula  $\eta=G\tau _{M}$ (the Maxwell relation) in analogy with 
 solids, where $G$ is the modulus of rigidity \cite {max}.
The macroscopic relaxation in Fig.4 is a result of the accumulation of 
microscopic relaxations by the excited particles. 
Quantum mechanics states that, in the decay from an excited state with an 
 energy $E$ to a ground state with $E_0$, the higher excitation 
 energy $E$ causes a shorter microscopic relaxation time $\tau $ as $\hbar /\tau ¥ \simeq |E-E_0|$.
  In eq.(8),  the left-hand side includes the $\tau _{M}$ 
in $\sigma _{1} =\rho /(4G\tau _{M})$, whereas  
 the right-hand side  includes the many-body excitation spectrum in $\chi^T(q,\omega)$.  
 In this sense,  eq.(8) is the many-body version of $\hbar /\tau ¥ \simeq |E-E_0|$.
 (In other words, eq.(8) is {\it another expression of the Maxwell relation\/}.)
In liquids, {\it slightly different microscopic
 local structures are irregularly arranged \/}, the energies $E$'s of which 
differ only slightly from each other. In the relaxation of one arrangement to others, the 
smaller energy difference $|E-E_0|$ leads to a longer $\tau $ in $\hbar /\tau ¥ 
\simeq |E-E_0|$. These long $\tau$'s result in 
a long $\tau _{M}$ of the macroscopic deformation, thus leading to a large 
shear viscosity $\eta =G\tau _{M}$ of the classical liquid.

In liquid helium 4 near $T_{\lambda}$, no structural transition is  
observed in coordinate space. Hence, $G$ must be a constant at the first 
approximation, and therefore the fall of the shear viscosity $\eta=G\tau _{M}$ is 
attributed to the decrease in $\tau _{M}$. At the microscopic level, 
this comes from the decrease in $\tau $'s. In view of 
$\hbar /\tau ¥ \simeq |E-E_0|$, the decrease in $\tau $  suggests the  
increase in $E$. Near $T_{\lambda}$, it is natural to attribute it 
to the effect of Bose statistics. The relationship between  
the excitation energy and  Bose statistics dates back to 
Feynman's argument on the scarcity of low-energy excitation in liquid helium 
4 \cite{fey1}, in which he explained  how Bose 
statistics affects the many-body wave function in configuration space. 
We will apply his explanation to the analysis of shear viscosity.

(2) Consider the motion of particles in Fig.4.  For example, two 
particles 1 and 2, each of which simultaneously starts at $(x,y)$ and 
$(x,y+\Delta y)$, move along the $x$-direction to positions 1' and 2'. 
The long thin arrows represent the displacement of white circles to black ones. 
In the BEC phase, the many-body wave function has permutation 
symmetry everywhere, and all white circles are therefore permutable in Fig.4.  
At first sight, these displacements by long arrows seem to be a large change,  
 but they are reproduced by a set of small displacements of 
 neighboring white circles by short thick arrows as shown in Fig.4. 
 In Bose statistics, owing to permutation symmetry, one cannot 
distinguish between two types of particles after displacement, one  moved 
from near positions by a short arrow, and the other moved 
from distant initial positions by a long arrow. {\it Even if  the 
displacement made by the long arrows is a large 
displacement in classical statistics, it is only a small 
displacement  by the short arrows in Bose statistics \/}. 
The displacement related to shear viscosity is a transverse 
one. In general, the transverse displacement does not change 
 particle density on a large scale, and therefore, for any given particle after displacement,   
a nearby particle always exists in the initial distribution.
(On the other hand, the longitudinal displacement largely changes the 
 particle density, and therefore nearby particles do not always exist, which 
implies $\chi^L(q,\omega)-\chi^T(q,\omega)¥\ne 0$ in eq.(9).)

Let us consider this situation in 3N-dimensional configuration space following Feynman. 
 The excited state related to shear viscosity, in which particles 
 make small displacements, has a wave function that is not far apart 
 from the ground-state wave function in configuration space.    
The excited-state wave function must be orthogonal to the 
ground-state wave function. Since the latter has an uniform amplitude,
 the former must spatially oscillate between the plus and minus values. 
This means that the wave function of the excited state due to slight displacements 
must oscillate within a small distance in configuration space. The kinetic energy of the system 
is determined by the 3N-dimensional gradient of the many-body wave 
function, and therefore  this steep rise and fall of amplitude raises excitation energy. 
The relaxation of such a microscopic state is a rapid process. When these 
processes occur simultaneously, it leads to a rapid relaxation of the 
macroscopic deformation. This mechanism intuitively explains why  Bose statistics leads  
to the decrease in shear viscosity $\eta =G\tau _{M}$.

(3) When the system is at high temperatures, the coherent wave function has a microscopic size.  
 If the long arrow in Fig.4 takes a particle out of such a wave function, 
 the particle after displacement cannot be reproduced by the slight displacement. 
 The mechanism below $T_{\lambda}$ does not work for such a displacement, 
 and $\tau _{M}$ changes to the ordinary long $\tau _{M}$ in the classical liquid.
When the system is in $T_{\lambda}< T < 3.7K$, the size of the coherent 
wave function is mesoscopic. In a repulsive system  with high 
density such as liquid, the large-distance motion takes much energy,  
and therefore in the low-energy excitation, particles are likely to stay 
within the same wave function. The  energy of such an excitation is low 
compared with that of the excitation due to large displacements,
but owing to Bose statistics, it is not so low as that in the 
classical liquid, and therefore its relaxation to the ground state is relatively a 
fast process \cite {sma}. As $T\rightarrow T_{\lambda}$,  
 the number of such fast relaxations  gradually increases, and 
 the superflow appears within the mesoscopic distance,
 which is the reason for the substantial fall of $\eta $ at $T_{\lambda} < T < 3.7K$.

\section{Microscopic Model}
\subsection {Onset of superflow} 
 To formulate the mechanism of the fall of $\nu $ above $T_{\lambda}$, 
 we consider the repulsive Bose system having the following hamiltonian 
\begin{equation}
 H=\sum_{p}\epsilon (p)\Phi_{p}^{\dagger}\Phi_{p}
   +U\sum_{p,p'}\sum_{q}\Phi_{p-q}^{\dagger}\Phi_{p'+q}^{\dagger}\Phi_{p'}\Phi_p , 
   \qquad (U>0),¥¥¥
	\label{¥}
\end{equation}¥
 where $\Phi_p $ is the annihilation operator of a spinless boson, and 
 begin with the state without the macroscopic condensate.  
 The onset of superflow above $T_{\lambda}$ has a similar 
mechanism to the onset of the nonclassical moment of inertia above $T_{\lambda}$. 
In this subsection, we recapitulate ref.14 on this point 
with some modifications.

  Let us discuss the role of particle interaction. The susceptibility
  $\chi_{\mu\nu}(q,\omega) $ is a correlation function of currents 
$j_{\mu}(q,\tau)=\sum_{p,n}(p+q/2))\Phi_{p}^{\dagger}\Phi_{p+q}e^{-i\omega _n\tau }$. 
From $\chi_{\mu\nu}=\chi^Lq_{\mu}q_{\nu}/q^2 +\chi^T 
(\delta_{\mu\nu}-q_{\mu}q_{\nu}/q^2)¥$,  we will extract  
 the term proportional to $q_{\mu}q_{\nu}$, and define 
 $\hat{\chi}_{\mu\nu}(q,\omega )=(\chi^L-\chi^T) q_{\mu}q_{\nu}/q^2$ for $\hat {\rho _s}(T)$.
 In the ideal Bose system,  we obtain
\begin{equation}
\hat{\chi}_{\mu\nu}(q,\omega)  
	            =-\frac{q_{\mu}q_{\nu}}{4¥}¥
	                    \frac{1}{V¥}\sum_{p}\frac{f_B(\epsilon (p))-f_B(\epsilon (p+q))}
	                                   {\omega+\epsilon (p)-\epsilon (p+q)¥}¥,
	\label{¥}
\end{equation}¥
where $f_B(\epsilon (p))$ is the Bose distribution.
If bosons would form the macroscopic condensate, $f_B(\epsilon (p))$ in eq.(14) is a macroscopic 
number for $p=0$ and nearly zero for $p\ne 0$. Thus, in the sum over $p$ on the 
right-hand side of eq.(14), only two terms corresponding to $p=0$ and 
$p=-q$ remain, with a result of $\hat{\chi}_{\mu\nu}(q,0)=\rho _s(T)¥q_{\mu}q_{\nu}/q^2¥¥$.
When bosons form no condensate, however, the sum over $p$ in 
eq.(14) is carried out by replacing it with an integral, and one notices 
that $q^{-2}$ dependence disappears, hence $\rho _s(T)=0$.
This means that when examining the system near $T_{\lambda}$, $\hat{\chi}_{\mu\nu}(q,\omega )$ 
 under the particle interaction $H_I$ is needed.  
 We obtain a perturbation expansion of $\chi_{\mu\nu}(q,\omega) $ with respect to $U$ as 
 \begin{eqnarray}
  \lefteqn{\langle G|T_{\tau}j_{\mu}(x,\tau)j_{\nu}(0,0)|G \rangle}   \\ 
       && =\frac{\displaystyle{\langle 0|T_{\tau}\hat{j}_{\mu}(x,\tau)\hat{j}_{\nu}(0,0)
 	              exp\left[-\int_{0}^{\beta¥}d\tau \hat{H}_I(\tau)¥\right]|0\rangle¥}}
 	        {\displaystyle{\langle 0|exp\left[-\int_{0}^{\beta¥}d\tau  \hat{H}_I(\tau)¥\right]|0\rangle¥}}¥.\nonumber
	\label{¥}	
\end{eqnarray}¥

\begin{figure}
\includegraphics [scale=0.5]{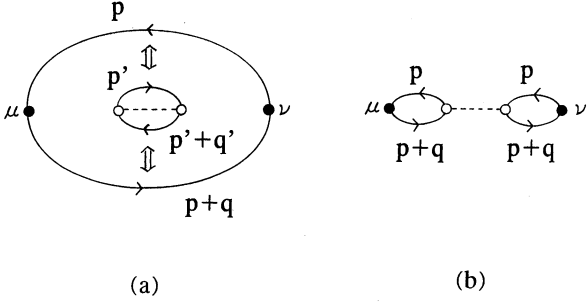}
\caption{\label{fig:epsart} When $p=p'$ and $p+q=p'+q'$ in (a), an 
          exchange of particles between large and small bubbles  yield (b). 
            }
\end{figure}

In eq.(8) and its modified form of $\sigma_{1n}(\omega)$ using $\chi^L(q,\omega)$, 
the characteristic excitations of a liquid is included in $\chi^L(q,\omega)$. 
$\hat{\chi}_{\mu\nu}(q,\omega )=(\chi^L-\chi^T) q_{\mu}q_{\nu}/q^2$
represents the change of $\chi^L$ induced by Bose statistics. 
When $\hat{\chi}_{\mu\nu}(q,\omega)$ first obtains a nonzero value at $T_{on}$, one can 
represent it by the simplest process in eq.(15).  Figure. 5(a) shows 
$j_{\mu}j_{\nu}$ as a large bubble, and the effect of $H_I$ as a small bubble with 
a dotted line $U$.  As $T\rightarrow T_{\lambda}$ in the normal 
phase, Bose statistics forces particles in the large and small bubbles 
to strictly obey the permutation symmetry. 
In Fig.5(a), when one of two momenta are equal ($p=p'$) for two bubbles, 
and when the other two momenta are also equal ($p+q=p'+q'$), a graph made 
by exchanging these particles must be included in the perturbation expansion. 
Cutting the line at the point denoted by opposite arrows in Fig.5(a), and reconnecting it 
to the line with the same momentum of the other bubble yields  
two bubbles in Fig.5(b) with the common momenta $p$ and $p+q$ connected 
by the repulsive interaction $U$. 
As the mesoscopic condensate grows, such an exchange occurs many  
times, thus leading to a large chain of bubbles. Furthermore, among various 
momenta, the process including particles with $p=0$ or $p+q=0$ grows to 
play a dominant role. As a result, we obtain 
\begin{eqnarray}
 \lefteqn {\hat{\chi}_{\mu\nu}(q,i\omega)} \\
	      && =\frac{q_{\mu}q_{\nu}}{4¥}¥
	      \frac{1}{V¥}\left[\frac{ F_{\beta}(q,i\omega)}{1-U F_{\beta}(q,i\omega)¥}¥
	            +\frac{F_{\beta}(q,-i\omega)}{1-U F_{\beta}(q,-i\omega)¥}¥ \right]¥,¥¥
		     \nonumber 
	\label{¥}	
\end{eqnarray}¥
where 
\begin{eqnarray}
	\lefteqn {F_{\beta}(q,i\omega) =}\\ 
	&&\frac{(\exp(\beta[\Sigma-\mu])-1)^{-1}-(\exp(\beta[\epsilon (q)+\Sigma-\mu])-1)^{-1}} 
	                         {-i\omega +\epsilon (q)¥¥} ¥.  \nonumber
	\label{}
\end{eqnarray}¥
($\Sigma $ is the self energy and $\mu $ is the chemical potential.) Since our interest is
$\chi^L -\chi^T$ near $\omega =0$, we expand eq.(16) with respect to $\omega$
\begin{eqnarray}
	\lefteqn {\hat{\chi}_{\mu\nu}(q,\omega) 
	       =\frac{q_{\mu}q_{\nu}}{4¥}¥ \frac{1}{V¥}
	             \frac{ 2F_{\beta}(q,0)}{1-U F_{\beta}(q,0)¥}¥}  \nonumber\\ 
		&& \times \left[1-\left(\frac{\omega}{\epsilon (q)¥(1-U F_{\beta}(q,0)¥)¥}\right)^2 
		                +\cdots ¥\right]¥. 
	\label{¥}
\end{eqnarray}¥

$F_{\beta}(q,0)$ in eq.(17) is a positive monotonically decreasing function of $q^2$. 
Since our interest is the macroscopic response of the system, let us 
expand it with respect to $q^2$ as 
\begin{eqnarray}
	\lefteqn{ F_{\beta}(q,0)=\frac{\beta}{4\sinh ^2 
	                      \displaystyle{\left(\frac{|\beta[\mu(T)-\Sigma]|}{2¥}\right)}¥¥¥}}\nonumber \\ 
	         && \times \left[1-\frac{\beta}{2¥}\frac{1}{\tanh  
	                   \displaystyle{\left(\frac{|\beta[\mu(T)-\Sigma]|}{2¥}¥\right)}¥¥}
	                        \frac{q^2}{2m¥}¥¥¥¥  +\cdots    \right]¥¥  \nonumber \\ 
	              &\equiv & a-bq^2+\cdots . 
	\label{¥}
\end{eqnarray}¥
The denominator on the right-hand side of eq.(18) has the form 
$1-UF_{\beta}(q,0)=(1-Ua)+Ubq^2+\cdots $. As $T$ approaches $T_{\lambda}$ 
($\mu(T)-\Sigma \rightarrow 0$), the first term $a$ in the 
expansion of $F_{\beta}(q,0)$ gradually increases. 
 Hence,  $1-Ua$ gradually decreases, and finally satisfies $1-Ua=0$.
At this temperature, eq.(18) becomes proportional to 
$q_{\mu}q_{\nu}/q^2$ as $V^{-1}[a/(2Ub)]q_{\mu}q_{\nu}/q^2$,
hence giving its coefficient $\chi^L -\chi^T$ a nonzero value.
This is {\it a sign of the onset of superflow by the mesoscopic condensate \/} \cite {koh}. 

\begin{figure}
\includegraphics [scale=0.4]{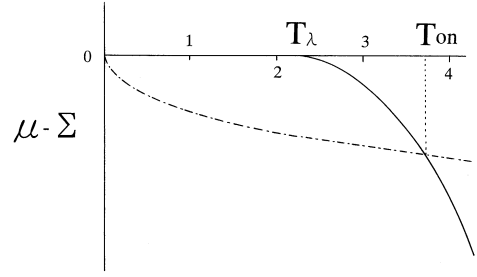}
\caption{\label{fig:epsart} The temperature dependence of $\mu 
(T)-\Sigma (U)$ depicted by  a solid curve, and the condition of 
eq.(20) by a one-point-dotted one. $T_{on}$ corresponds to 3,7K in Fig.1. }
\end{figure} 

The condition $1-Ua=0$ yields
\begin{equation}
     U\beta=4\sinh ^2\left(\frac{\beta[\mu (T)-\Sigma]}{2¥}¥\right)¥ .
	\label{¥}
\end{equation}¥
Figure. 6 schematically shows the condition of eq.(20) by a one-point-dotted 
curve, and $\mu (T)-\Sigma$ of the system by a thick curve. (The 
simplest approximation for $\mu (T)$ of liquid helium 4 is to use $\mu (T)$ of the ideal Bose gas 
with $T_c$ replaced with $T_{\lambda}$,
\begin{equation}
   	 \mu (T)-\Sigma (U)=-\left(\frac{g_{3/2}(1)}{2\sqrt{\pi}¥}¥\right)^2k_BT_{\lambda}
	          \left[\left(\frac{T}{T_{\lambda}¥}\right)^{3/2}-1\right]^2,
  	        \label{¥}
\end{equation}¥ 
which approximation dates back to London's paper.)  Since $\mu (T)-\Sigma$ reaches zero 
at the finite temperature $T_{\lambda}$, it approaches zero more 
rapidly than the one-point-dotted curve by eq.(20). Hence, above $T_{\lambda}$, 
there is always a temperature in which eq.(20) is satisfied. This is the reason why 
$\nu (T)$ in eq.(11) begins to fall above $T_{\lambda}$ as in Fig.1. We call 
this temperature {\it the onset temperature \/} $T_{on}$. Using the definition of $a$ and $b$ in 
eq.(19) and $Ua=1$ in $V^{-1}[a/(2Ub)]q_{\mu}q_{\nu}/q^2$, the 
mesoscopic superfluid density
$\hat {\rho _s}(T)=\lim_{q\rightarrow 0}¥[\chi^L -\chi^T]$ at $T_{on}$ has the form  
\begin{equation}
	\hat {\rho _s}(T)=\frac{1}{V¥}\frac{m}{\sinh |\beta [\mu(T)-\Sigma] |¥}¥.
\end{equation}¥
Since the sizes of the condensate and capillary are not so different, we 
do not take the $V \rightarrow \infty $ limit in eq.(22). Hence even when 
$\mu -\Sigma \ne 0$, $\hat {\rho _s}(T)$ is finite. Using eq.(22) in  
eq.(11), we can describe the  fall of $\nu (T)$ in Fig.1.
Equation (22) serves as an interpolation formula of  $\hat {\rho _s}(T)$ in 
$T_{\lambda}<T<3.7K$, and jumps to a macroscopic value at $T_{\lambda}$.

\subsection{Conductivity spectrum} 
Let us obtain the expression of $\sigma _1(\omega)$ in eq.(9). 
In  Fig.2(b), the emergence of the peak at $\omega =0$ is 
accompanied by a gradual change of the continuum part. 
At $T=T_{on}$ ($Ua=1$), we rewrite eq.(18) as
 \begin{equation}
		\hat{\chi}_{\mu\nu}(q,\omega)=\frac{q_{\mu}q_{\nu}}{q^2¥}¥\frac{a^2}{2Vb¥}¥
		 \left[1-\frac{\omega ^2}{\omega _d^2¥}+ \cdots 	¥\right]¥,
	\label{¥}
\end{equation}¥
where 
\begin{equation}
	 \omega _d^2=\frac{\epsilon (q_{e})^4}{4(k_BT)^2¥}¥\frac{1}{\tanh ^2 
	 \displaystyle{\left(\frac{|\beta[\mu(T)-\Sigma]|}{2¥}¥\right)}¥}¥.
	\label{¥}
\end{equation}¥
($q_e$ is the wave number representing the size of the capillary.)
 We want a simple formula of $\hat{\chi}_{\mu\nu}(q,\omega)$ that  
 agrees with eq.(23) at $\omega \rightarrow 0$ and shows a reasonable behavior
at $\omega \rightarrow \infty $.  
 At $T_{on}$, the simplest form of $\hat{\chi}_{\mu\nu}=(\chi^L-\chi^T) q_{\mu}q_{\nu}/q^2$
 satisfying these conditions is given by
 \begin{equation}
	\hat{\chi}_{\mu\nu}(q,\omega)
	     \simeq \frac{q_{\mu}q_{\nu}}{q^2¥}¥\frac{a^2}{2Vb¥}¥
	            \frac{1}{\left(¥\displaystyle{1+\frac{\omega ^2}{\omega _d^2¥}}¥\right)¥}¥.
	\label{¥}
\end{equation}¥
 Equation.(25) represents the dynamical balance between the longitudinal 
and transverse responses. Using the definitions of $a$ and $b$ in 
eq.(19) and $Ua=1$, we define the generalized form of the 
mesoscopic superfluid density $ \hat{\rho _s}(T,\omega)\equiv
\lim _{q\rightarrow 0}[\chi^L(q,\omega)-\chi^T(q,\omega)]$ as 
\begin{equation}
  	    \hat{\rho _s}(T,\omega)¥=\frac{1}{V¥}¥ \frac{m}{\sinh |\beta [\mu(T)-\Sigma] |¥}¥ 
		 \frac{1}{\left(¥\displaystyle{1+\frac{\omega ^2}{\omega _d^2¥}¥}¥\right)}. 
	\label{¥}
\end{equation}¥
 Using eq.(26) as $\chi^L-\chi^T$ in eq.(9), taking the finite part of the integral as 
\begin{equation}
	\int_{0}^{\infty ¥}¥d\omega\frac{1}{\omega ^2-\omega '^2¥}¥
	\frac{1}{\left(¥\displaystyle{1+\frac{\omega ^2}{\omega _d^2¥}¥}¥\right)}¥
	   = \frac{-\pi}{2\omega _d\left(¥\displaystyle{1+\frac{\omega '^2}{\omega _d^2¥}¥}¥\right)}¥¥,
	\label{¥}
\end{equation}¥
and eq.(24) for $\omega _d$, we obtain the real part of conductivity as
\begin{eqnarray}
 \lefteqn{ \sigma _1(\omega)=\sigma_{1n}(\omega) 
           +\frac{2V^{-1}}{\rho\pi 
		  d^2¥}\frac{m}{\sinh |\beta [\mu(T)-\Sigma] 
		  |¥}\frac{\pi ^2¥}{2}¥\delta (\omega ) } \\  
	 &&-\frac{2V^{-1}}{\rho\pi d^2¥}¥\left(\frac{k_BT}{\epsilon (q_e)^2¥}\right)¥¥
                    \frac{m}{ \displaystyle{\cosh ^2 \left(\frac{|\beta [\mu(T)-\Sigma] |}{2¥}\right)¥¥}¥}¥
           \frac{\pi}{\left(1+\displaystyle{\frac{\omega ^2}{\omega _d^2¥}¥}¥\right)}¥  . \nonumber
	\label{¥}
\end{eqnarray}¥
 $\sigma _1(\omega)$ must satisfy the sum rule eq.(5) whether it is in the normal or superfluid phases.  
(In $\int_{0}^{\infty¥}¥\sigma _1(\omega)d\omega$ of the sum rule,  the second term 
in  the bracket of the right-hand side of eq.(28) yields a 
term proportional to $1/\sinh |\beta [\mu(T)-\Sigma]| $ with the aid of
$\int_{0}^{\infty¥}¥d\omega/(1+\omega^2/\omega _d^2)=\pi \omega _d/2$ and eq.(24). 
The first and second terms in the bracket of eq.(28) cancel out each other 
in $\int_{0}^{\infty¥}¥\sigma _1(\omega)d\omega$. \cite {del})

 In eq.(28), $\sigma _{1n}(\omega )$ is given by the real part of eq.(B5) 
in Appendix. B. Hence, the conductivity of the normal-fluid part $\sigma _n(\omega )$ is given by
\begin{eqnarray}
  \lefteqn{\sigma _n(\omega )=\frac{1}{\omega d^2¥}Im\left(1-\frac{1}{J_0\left(id[1+i]
                           \displaystyle {\sqrt {\frac{\omega}{2\nu _n¥}¥}}\right)¥¥}\right)¥}  \\ 
              && -\frac{2}{nd^2¥}¥\left(\frac{k_BT}{\epsilon(q_e)^2¥}\right)¥¥
                    \frac{V^{-1}}{ \displaystyle{\cosh ^2\left(\frac{|\beta [\mu(T)-\Sigma] |}{2¥}¥\right)¥ }¥}¥
                    \frac{1}{\left(1+\displaystyle{\frac{\omega 
		    ^2}{\omega _d^2¥}¥}¥\right)}¥,\nonumber
	\label{¥}
\end{eqnarray}¥
($J_0$ is the zeroth order Bessel function, and $n=\rho /m$).
The temperature dependence of $ \sigma _n(\omega )$ comes from the 
second term on the right-hand side  
of eq.(29).  As $\mu(T)-\Sigma \rightarrow 0$, $1/\cosh ^2 |\beta 
[\mu(T)-\Sigma]/2|$ and $\omega _d$ increase, hence, 
$\int_{0}^{\infty¥}¥ \sigma _n(\omega )d\omega$ decreases.  
On the other hand, the conductivity of the superfluid part $\sigma _s(\omega )$ satisfies
\begin{equation}
	\int_{0}^{\infty ¥}¥\sigma _s(\omega )d\omega=\frac{\pi }{2nd^2¥}¥
	     \frac{V^{-1}}{\sinh |\beta [\mu(T)-\Sigma] |¥}¥ ¥.
	\label{¥}
\end{equation}¥
As $\mu(T)-\Sigma \rightarrow 0$, the sharp peak at $\omega =0$ increases owing to 
$1/\sinh |\beta [\mu(T)-\Sigma] |$. Equation (29) and (30) are the 
simplest formulae describing the change in $\sigma (\omega)$ at
$T_{\lambda}<T<3.7K$ in Fig.2. 

\subsection{Stability of superflow} 
  Equation. (26) implies that, for the objects oscillating at the frequency above $\omega _d$,  
a considerable part of the superfluid behaves as a normal flow. 
Here, we call $\omega _d$ in eq.(24) {\it the damping angular frequency \/}. 
For the experiment using the vertically standing capillary in gravity, 
we measure the integrated response of the system from $\omega =0$ to $\omega _f$. 
As long as the properties at approximately $\omega =0$ 
determine the observed quantities, this feature above $\omega _d$ does not 
seriously affect the result. On the other hand, there is another type 
of experiment in which the superfluid behavior is measured only at  certain frequencies,
such as torsional oscillation or ultrasound. In this method, there is a case in which the above 
feature gives rise to a serious problem (see \S 4).

The stability of the superflow by the mesoscopic condensate depends on temperature.
As the Bose statistical coherence grows in eq.(24) ($\mu(T)-\Sigma \rightarrow  
0$), the superflow becomes robust to the oscillating probe in eq.(26) 
($\omega _d \rightarrow \infty $). 
By gradually changing temperature at $T>T_{\lambda}$, we can 
realize different size distributions of the mesoscopic condensate in the capillary flow.  
The quantity reflecting the size of the condensate is the chemical 
potential $\mu (T)$. Here, to characterize the size distribution of the
 condensate at $T$,  we use the number of particles $\langle s\rangle$ satisfying 
 $\beta (\mu -\Sigma)\langle s\rangle=-1$ (see Appendix.C).
For a small $\mu -\Sigma$, eq.(24) is rewritten as 
\begin{equation}
	\omega _d  =¥\frac{\epsilon (q_{e})^2}{k_BT¥}¥\langle s\rangle.
	\label{¥}
\end{equation}¥
 As $\langle s\rangle$ grows, $\omega _d$ increases, hence, as the 
size of the mesoscopic condensate grows, the superflow by the  
mesoscopic condensate gradually becomes robust to the oscillating probe.

The stability of superflow is also related to the repulsive interaction 
$U$. In the famous argument by Landau on the critical velocity of fully developed BEC, 
the change of one-particle spectrum from $p^2/(2m)$ to $v_sp$ induced by the 
repulsive interaction in the Bogoliubov spectrum plays a crucial role \cite {lan}. 
Although Landau's argument deals with the magnitude of critical 
 velocity, eq.(26) deals with the upper limit of frequency at which flow velocity changes.
When $U=0$ in eq.(18), $\hat{\rho _s}(T,\omega)$ is written in the 
form of eq.(26), with $\sinh |\beta [\mu(T)-\Sigma] |$ replaced by 
$\sinh ^2(|\beta [\mu(T)-\Sigma] |/2)$. In this case, $\omega _d^2$ in 
eq.(24) is simply $\epsilon (q_e)^2$. Hence,  
 with increasing $\omega$,  $\hat{\rho _s}(T,\omega)$ vanishes 
 far more  rapidly than that in the case of $U\ne 0$. 
Physically, the repulsive interaction $U$ prevents the drop of particles from the superflow, 
thus stabilizing it. In eq.(24), the repulsive interaction gives 
the factor $1/\tanh ^2(|\beta[\mu(T)-\Sigma]|/2)$ to $\omega _d^2$.
Hence, $\hat{\rho _s}(T,\omega)$ does
not easily vanish for a large $\omega$ in eq.(26). 
 {\it The repulsive interaction plays the significant role not only in 
the emergence of superflow by mesoscopic condensate, but also in 
its stabilization.\/}  In this sense, eqs.(24) and (26) are extensions 
of Landau's argument to the underdeveloped BEC.

\section{Implication for Superflow in Porous Media} 
In the capillary flow above $T_{\lambda}$, the  mesoscopic condensate  spontaneously 
emerges in free space, whereas in porous media below $T_c$, it 
is forced to emerge in a restricted space \cite {tak}.  Despite this difference, 
when we focus on a  single flow through a pore, the result of this paper gives
us a simple criterion for the stability of superflow in porous media. 
In porous media, the superflow is normally measured by the method of 
torsional oscillation or ultrasound. Just below $T_c$, $\langle s \rangle$ in eq.(31) 
and $\omega _d$ in eq.(26) are still small. When the oscillator at $\omega >\omega  _d$ induces 
 the oscillation of superfluid, $[\hat{\rho  _s}(T,0)-\hat{\rho _s}(T,\omega)]/\hat{\rho _s}(T,0)$  
 of the superfluid behaves as a normal-fluid part. Hence,  {\it $\omega ^2/(\omega ^2+\omega_d^2)$ 
 of superfluid is locked to the substrate \/}, and $\hat {\rho _s}(T)$ is underestimated.
With decreasing temperature, this underestimation of $\hat {\rho _s}(T)$ is 
improved by an increase in $\omega _d$  \cite {ind}. 
Alternatively, when we change the frequency at a given temperature,
 the oscillation methods using a higher frequency will detect a larger discrepancy between  
 $T_c$ in the mechanical response and  $T_c$ in the thermodynamic one. 

On the other hand, in the oscillation experiments on the  
superfluid, another type of upper limit $\omega _p$ is known,  
 which is set for the normal-fluid part to be locked to the substrate. (The viscous 
penetration depth $\delta =\sqrt {\nu /(2\omega)}$ must be greater than 
the channel size. Since the relevant  size of the channel is about 
$10^{-5}cm$ and $\nu \simeq 10^{-4}cm^2/s$, $f_p=\omega _p/2\pi $ is roughly $10^6$ Hz.) 
In normal situations, $\omega _p > \omega _d$ is usually satisfied.  
At $\omega >\omega _p$, neither the normal-fluid nor superfluid 
parts is locked to the substrate, hence, $\hat {\rho _s}(T)$ is overestimated. 
At $\omega _d>\omega $, the normal and superfluid parts are correctly discriminated. 
A delicate problem arises at $\omega _p>\omega >\omega _d$, 
especially when a frequency comparable to $f_p=\omega _p/2\pi $ is used in ultrasound. 
Although $\omega _p$ does not depend on temperature, $\omega _d$ 
increases with decreasing temperature, and there may be a case of  
$\omega _d > \omega _p$ at very low temperatures.
When we represent $\omega _d$ by its maximum $\omega _d(T=0)$, we can 
classify the situation into two cases: $\omega _p >\omega _d(T=0) $ 
and $\omega _d(T=0) >\omega _p$, depending on the type of porous medium. 
The interpretation of the experimental data obtained in such a situation will be 
complicated, because of the existence of two different kinds of upper limit in frequency.

Recently, a comparison between the measurement using a torsional oscillator 
(2140 Hz) and by ultrasound (10 MHz) was made in liquid helium 4 
confined in a porous substrate, hectorite, and Gelsil \cite {tan}. 
The decoupling of the superfluid 
from the oscillating substrate results in a decrease in the moment of inertia $\Delta I_z$, 
and in the increase in the  ultrasound velocity $\Delta v_s$. 
Different situations will be realized according to different 
frequencies and substrates.  The comparison of experimental data  
between different frequencies will give us a clue to the  
dynamics of superflow in porous media.

\section{Discussion}

 \subsection{Comparison between superflow in the dissipative process 
 and that in the nondissipative one}   
On the onset mechanism of superflow, it makes a difference whether 
superflow appears in the dissipative or nondissipative process. Comparing 
the moment of inertia $I_z(T)= I_z^{cl}(1-\hat {\rho _s}(T)/\rho)¥$ 
for the  non-dissipative process  \cite {koh} with the kinematical viscosity 
$\nu (T)$ of eq.(11) for the dissipative one, we note the following features of $\nu (T)$. 

(1) In $I_z(T)= I_z^{cl} (1-\hat {\rho _s}(T)/\rho)¥$, $\hat {\rho _s}(T)$ appears 
only in the coefficient of the linear term of $I_z^{cl}$. In $\nu (T)$ 
of eq.(11), the effect of Bose statistics 
appears in all coefficients of higher-order terms of $\nu _n$ except for the first-order one. 
This feature does not depend on the particular model of a  
liquid, but on the general argument. (On the other hand, the microscopic 
derivation of $\nu _n$ depends on the model of a liquid.) 

(2) In $I_z(T)= I_z^{cl} (1-\hat {\rho _s}(T)/\rho)¥$, the change in 
$\chi ^L(q,\omega )-\chi ^T(q,\omega )$ directly affects $I_z$ without being enhanced,  
 and therefore the observed effect of the finite $\hat {\rho _s}(T)$ 
 above $T_{\lambda}$ is very small. 
 In $\nu (T)$ of eq.(11), because of the sharp peak in the dispersion integral in eq.(9),
the small change in $\hat {\rho _s}(T)$ is strongly enhanced to the 
 observable change in $\nu (T)$. 

(3) The existence of $1/d^2$ before $\hat {\rho _s}(T)/\rho$ in eq.(11) indicates 
that the narrower capillary shows clearer evidence of frictionless flow.  
 Similarly, the existence of $1/\omega _f$ indicates that the choice of 
 experimental procedure, such as the method of applying the pressure 
 between two ends of the capillary, affects the temperature 
 dependence of the $\nu (T)$ of capillary flow.
This means that {\it superflow appearing in the dissipative 
process depends on more variables than that in the non-dissipative one\/}, 
which is in accordance with the general feature of nonequilibrium phenomena.

\subsection{Comparison with thermal conductivity}   
 In liquid helium 4 near $T_{\lambda}$, we know a marked change in another type  
of conductivity, the anomalous thermal conductivity. 
 Under a given temperature gradient $\nabla T$, the heat flow $Q$ satisfies 
 $Q=-\kappa\nabla T$, where $\kappa$ is the coefficient of thermal conductivity. 
 In the critical region above $T_{\lambda}$ ($|T/T_{\lambda}-1|<10^{-3}$), 
 the rapid rise of $\kappa$ is  observed, and finally at $T=T_{\lambda}$, 
 $\kappa$ abruptly jumps to infinity.  For $T_{\lambda}<T<3.7K$, 
 however, two kinds of conductivity behave differently. In shear 
 viscosity, the corresponding conductivity $\sigma (T)=1/(4\nu)$ shows  
 a gradual rise, whereas $\kappa (T)$ shows no such rise. 
This difference is expected for the following reason.
The $\kappa $ in thermal conductivity is expressed in terms of 
the  correlation function with a similar structure 
 to the $\eta $ in shear viscosity, but $\kappa$ and $\eta$ are qualitatively different.
 Although shear viscosity is associated with the transport of momentum (a vector), 
 thermal conductivity is associated with that of energy (a scalar).  For the vector field (velocity 
 field), the direction of vectors has a rich variety in its spatial 
 distribution.  Among various flow-velocity fields, we  can   
 regard the Couette flow rotating like a rigid body in a bucket 
 as the nondissipative counterpart to the  capillary flow.  
 On the other hand, for the scalar field (temperature field), the variety 
 of possible spatial distributions is far limited. The flow of heat energy is 
 always a dissipative phenomenon, and there is no  nondissipative counterpart.  
 Hence, the formulation using the Kramers-Kronig relation in \S 2 cannot be 
 applied to the onset of the anomalous thermal conductivity, 
 and no mechanism that amplifies the small $\hat {\rho _s}(T)$ is expected. 
 This formal difference between shear viscosity  and thermal conductivity 
  is consistent with the experimental difference between $\sigma (T)=1/(4\nu)$ and 
  $\kappa (T)$ at $T_{\lambda}<T<3.7K$.

\subsection{Comparison with  Fermi liquids}   
 The fall of the shear viscosity in liquid helium 3 at $T_c$ is  
 a phenomenon  parallel to that in liquid helium 4.  The formalism 
in \S 2 is applicable to liquid helium 3 as well.  For the behavior above $T_c$, however, 
there is a striking  difference between   liquid helium 3 and 4. 
The phenomenon occurring in fermions in the vicinity of $T_c$ is not the 
gradual growth of the coherent wave function, but the formation of 
Cooper pairs by two fermions.  
 (This difference evidently appears in the temperature dependence of  
 specific heat:  The $C(T)$ of  liquid helium 3 shows a sharp peak at $T_c$ 
without a symptom of its rise above $T_c$.)
 Once Cooper pairs formed, they are  composite bosons with a high 
 density at low temperatures,  and immediately jump to the superfluid state. 
Hence, the shear viscosity of liquid helium 3 shows an abrupt drop at 
$T_c$ without a gradual fall above $T_c$, like the dotted line in Fig.1.

\subsection{Future problems} 
As $T$ approaches $T_{\lambda}$, eq.(22) becomes insufficient to use as 
an interpolation formula of $\hat {\rho _s}(T)$, because it predicts a far smaller 
 $\hat {\rho _s}/\rho$ than that in Fig.3  when eq.(21) is used. 
 This implies that a considerable number of helium 4 atoms with $p \ne 0$ are dragged 
into the superflow.  Owing to  the repulsive interaction, 
particles are likely to spread uniformly in coordinate space.  By this feature,  
 particles with $p\ne 0 $ are forced to behave similarly to particles with $p=0$. 
 {\it If they move differently from the superflow, 
 particle density becomes locally high, thus increasing interaction energy.\/}  This is the 
reason why {\it a considerable number of particles with $p \ne 0$ participate in the 
superflow even when it is above $T_{\lambda}$ \/}. This tendency will 
become apparent as $T\rightarrow T_{\lambda}$.  Hence, the observed 
$\hat{\rho _s}(T)$ may be the sum of 
all $\hat{\rho _s}(T)$'s over different momenta. To explain Fig.3 
quantitatively, a more realistic model of liquids is needed.

Among all nonequilibrium phenomena in liquids, the capillary flow of
liquid helium 4 takes a unique position.
When increasing temperature from below to above $T_{\lambda}$, 
one can see a primitive form of the shear viscosity of liquids in the 
gradual rise in $\nu (T)$, which arises from the superfluid in such a way 
that it does not occur in other systems. The emergence of slightly 
different local structures leads to a long relaxation time $\tau _M$, 
and hence to a finite shear viscosity of liquids. Its study  
 has a possibility of shedding new light on the liquid theory. 

In this paper, we used the Poiseuille solution of classical fluid 
dynamics, and examined the quantum effect appearing in the coefficient of shear 
viscosity. As an alternative approach, we know the quantum 
hydrodynamics by Landau  \cite {lan} \cite {Miy}. If we consider the shear 
viscosity at $T_{\lambda}<T<3.7K$ using this approach, it will 
correspond to the semiclassical solution in quantum hydrodynamics.
This will give us interesting problems from practical and formal viewpoints.

\appendix

\section{$h(t)=h(0)\exp (-\alpha t)$}
To measure $\eta $, let us set the capillary to stand vertically in gravity $g$, 
the upper end of which is connected to a reservoir, while the lower 
end is open to the helium bath (Fig.1 in ref. 8).
The mass of liquid passing through the capillary per unit time is 
$Q=2\pi \rho \int_{0}^{d¥}¥rv_x(r)dr$, which is given by eq.(1) as  
\begin{equation}
   Q=\rho\frac{\pi d^4}{8\eta¥}¥\frac{\Delta P}{L¥}¥.
\end{equation}¥
$h(t)$ is the difference between the level inside the reservoir and the level in the bath.
In the reservoir with a radius $R$, a mass of liquid $\rho\pi R^2h(t)$  
flows out through the capillary at a rate of $Q$. 
In the vertically standing capillary, $\Delta P(t)$ between the above two levels  
 is $\rho gh(t)$ in eq.(A1). $h(t)$ decreases according to 
\begin{equation}
  \rho\pi R^2\frac{dh}{dt¥}¥ =-\frac{\pi d^4}{8\eta¥}¥\frac{\rho ^2 g}{L¥}¥h,
\end{equation}¥
thereby leading to $h(t)=h(0)\exp (-\alpha t)$ with $\alpha =\rho gd^4/(8\eta LR^2)$.

 \section{Conductivity Spectrum $\sigma (\omega)$}
The Stokes equation under the oscillating pressure gradient $\Delta P \exp (i\omega t)/L$ 
is written in the cylindrical polar coordinate as 
 \begin{equation}
 \frac{\partial v}{\partial t¥}¥=\nu \left( \frac{\partial }{\partial r^2¥}+ \frac{\partial }{r\partial r¥}\right)v
                              + \frac{\Delta P \exp (i\omega t)}{\rho¥L}¥.
\end{equation}¥
Velocity has the form
\begin{equation}
 v(r,t)¥=\frac{\Delta P\exp (i\omega t)}{i\omega \rho¥L}+\Delta v(r,t)¥,
\end{equation}¥
under the boundary condition $v(d,t)=0$. In eq.(B1), $\Delta v(r,t)$ satisfies 
\begin{equation}
 \frac{\partial \Delta v(r,t)}{\partial t¥}¥=\nu \left( \frac{\partial}{\partial r^2¥}
                      + \frac{\partial }{r\partial r¥}\right) \Delta v(r,t)¥,
\end{equation}¥
which solution is written in terms of the Bessel function $J_0(i\lambda r)$
with $\lambda =(1+i)\sqrt {\omega /(2\nu)}$. Hence,
\begin{equation}
 v(r,t)=\frac{\Delta P\exp (i\omega t)}{i\omega \rho¥L¥}¥
                       \left(1-\frac{J_0(i\lambda r)}{J_0(i\lambda d)¥}\right)¥.
\end{equation}¥
Using $v(r=0,t)$ of eq.(B4), the conductivity spectrum $\sigma (\omega )$ 
satisfying $\rho v(0,t)=\sigma (\omega )d^2 \Delta P\exp (i\omega t)/L$ (eq.(2)) is given by 
\begin{equation}
 \sigma (\omega )=\frac{1}{i\omega d^2¥}\left(1 
 -\frac{1}{J_0\left(id[1+i]\displaystyle {\sqrt {\frac{\omega}{2\nu¥}¥}}\right)¥¥}\right)¥.
\end{equation}¥
The real part of eq.(B5) gives a curve of $\sigma (\omega)$ in Fig.2(a).  
(Re $\sigma (0)$ in eq.(B5) agrees with $\rho /(4\eta )$.)  
It also determines the conserved quantity 
$\pi ^{-1} \int \sigma (\omega)d\omega =f(d)\propto d^{-2}$ in eq.(5).

 \section{Size of the Condensate}
The grand partition function of the ideal Bose system 
$Z_0(\mu)=\prod(1-\exp [\beta(\epsilon_p-\mu)]¥)^{-1}$ was rewritten by Feynman and Matsubara 
in terms of the size distribution of the coherent wave function \cite {fey2}. 
For the repulsive Bose system with $\mu (T)-\Sigma $, it is given by
\begin{eqnarray}
      \lefteqn{Z(\mu)=}\\
            &&\exp \left[\sum_{s=1}^{\infty ¥}
	             \left(\frac{\exp [\beta (\mu -\Sigma)s]}{s¥}
	         + A_s\frac{V}{\lambda ^3} \frac{\exp 
		 [\beta(\mu-\Sigma)s]}{s^{5/2}¥} 
		 \right)¥\right]¥,\nonumber 
	\label{¥}
\end{eqnarray}¥

where $s$ is the number of Bose particles participating in the 
coherent many-body wave function, $\lambda $ is the thermal wavelength, 
and $A_s$ is a weakly $s$-dependent factor.  The first and 
second terms in the exponent of eq.(C1) come  from $p=0$ and $p\ne 0$ bosons, 
respectively. In $\exp [\beta (\mu -\Sigma)s]/s$ of the first term, $1/s$ is 
a symmetry factor, and $\exp [\beta (\mu -\Sigma)s]$ represents the 
probability that the condensate has $s$ particles. 
In $\exp (-\beta |\mu -\Sigma|s)$, the particle number $\langle s\rangle$ satisfying 
$\beta (\mu -\Sigma)\langle s\rangle=-1$ gives us a rough estimate of 
the typical volume size of the condensate.


\begin{thebibliography}{99}
\bibitem{ide} The ideal Bose gas shows infinite fluctuations of the number of particles
		 at $T<T_c$. But one cannot naively expect that the actual interacting Bose 
		 systems also show large fluctuations in the especially wide range of temperature
		 near $T_{\lambda}$.
\bibitem{sma} In Fig.1, $\nu (3.7K)$ is already $10^{-3}$ times smaller than 
		   the $\nu $ of an ordinary classical liquid.
\bibitem{inc}  If $\nu (T)$ increases in the vicinity of $T_{\lambda}$, 
           thermal fluctuation seems to be a natural interpretation, 
           because the critical slowing down is likely to enhance the viscosity.
\bibitem{dis}  For the rigid-body rotation  
                $\mbox{\boldmath $v$}_d( \mbox{\boldmath $r$})= \mbox{\boldmath 
            $\Omega$}\times \mbox{\boldmath $r$}$, the dissipation function 
           $\Phi (\mbox{\boldmath $r$})=2\eta (e_{ij} - e_{kk}\delta_{ij}/3)^2¥$ 
	   is zero at every $r$,
	 where $2e_{ij}=\partial v_i/\partial x_j+\partial v_j/\partial x_i$. 
	 Although this flow is unstable due to the absence of an energy barrier 
	 in an actual viscous liquid, it is intrinsically nondissipative in an ideal circumstance.
\bibitem{ind} In practice, there remains a locked fraction 
	 	 of superfluid to substrates, which is independent of temperature and 
		frequency, and is relevant to internal geometries of cells.
\end{thebibliography}
\end{document}